\documentclass[sigconf]{acmart}
\AtBeginDocument{%
  \providecommand\BibTeX{{%
    \normalfont B\kern-0.5em{\scshape i\kern-0.25em b}\kern-0.8em\TeX}}}

\setcopyright{acmcopyright}
\copyrightyear{2023}
\acmYear{2023}
\acmDOI{XXXXXXX.XXXXXXX}

\acmConference[WSDM Cup'23]{Make sure to enter the correct
  conference title from your rights confirmation emai}{February 27-March 3,
  2023}{Singapore, Singapore}
\acmPrice{15.00}
\acmISBN{978-1-4503-XXXX-X/18/06}

\begin{document}

% \title{Ensemble-based Pre-training Model with Efficient Augmentation for Web Search}
% bert -> Pretrains -> E
% bert -> fintune -> E

% \title{Ensemble-enhanced Ranking Model with Multi-Pretraining Strategies for Web Search}
\title{Ensemble Ranking Model with Multiple Pretraining Strategies for Web Search}
% %%
% %% The "author" command and its associated commands are used to define
% %% the authors and their affiliations.
% %% Of note is the shared affiliation of the first two authors, and the
% %% "authornote" and "authornotemark" commands
% %% used to denote shared contribution to the research.

\author{Xiaojie Sun} % $^{\ast}$}
\authornotemark[1]
% \thanks{* All authors contribute equally, the order is determined randomly.}
\thanks{*Equal contribution, and the order is determined randomly.}

\affiliation{
 \institution{CAS Key Lab of Network Data Science and Technology, ICT, CAS}
 \institution{University of Chinese Academy of Sciences}
 \city{Beijing}
 \country{China}
}
\email{sunxiaojie21s@ict.ac.cn}

\author{Lulu Yu} % $^{\ast}$}
\authornotemark[1]
\affiliation{
 \institution{CAS Key Lab of Network Data Science and Technology, ICT, CAS}
 \institution{University of Chinese Academy of Sciences}
 \city{Beijing}
 \country{China}
}
\email{nothing_0_1@163.com}

\author{Yiting Wang} %$^{\ast}$}
\authornotemark[1]
\affiliation{
 \institution{CAS Key Lab of Network Data Science and Technology, ICT, CAS}
 \institution{University of Chinese Academy of Sciences}
 \city{Beijing}
 \country{China}
}
\email{wangyiting21s@ict.ac.cn}

\author{Keping Bi}
\affiliation{
 \institution{CAS Key Lab of Network Data Science and Technology, ICT, CAS}
 \institution{University of Chinese Academy of Sciences}
 \city{Beijing}
 \country{China}
}
\email{bikeping@ict.ac.cn}

\author{Jiafeng Guo}
\affiliation{
 \institution{CAS Key Lab of Network Data Science and Technology, ICT, CAS}
 \institution{University of Chinese Academy of Sciences}
 \city{Beijing}
 \country{China}
}
\email{guojiafeng@ict.ac.cn}

% \blfootnote{$^{*}$ All authors contribute equally, the order is determined randomly.}

% \affiliation{%

\begin{abstract}

  % How to effectively use the large-scale biased and noisy user feedback data together with relevance labels to the pre-training model is a promising topic. In this paper, we present our solution on the task of pre-training for web search in WSDM Cup 2023. 
  %不要放一句话来写，看起来太复杂了。先介绍有价值，再介绍limitations.
  % Using effective representation of relevance to train a pre-training model.
  % Pre-training models plays
  % The ranking model plays an important role in the retrieval system, but it requires a large amount of training data that can effectively represent the relevance between documents and queries. 
 An effective ranking model usually requires a large amount of training data to learn the relevance between documents and queries. 
 % Large-scale implicit user feedback (e.g., user click) has high mining potential, but it is also accompanied with bias and noise. How to make full use of these biased data to get an effective ranking model is a promising topic. In this paper, we propose an ensemble-enhanced ranking model. Firstly, multiple strategies such as incorporating additional negative samples and changing the objective via a novel correction method are used to obtain different pre-trained models. Further, we analyze the characteristic of the annotation set, adopt an efficient augmentation strategy and utilize an improved pairwise loss function to boost the performance. The Gradient Boosted Decision Tree was used for the ensemble at last. Our approach obtained competitive results and won the $3^{rd}$ place at task2.
  % We used Transformer with different layers as the main model, combined with different propensity weighting techniques for training. Specifically, we incorporated additional negative samples and changed the objective via a novel correction method. During the fine-tuned stage, we analyze the characteristic of the annotation set, adopt an efficient augmentation strategy and utilize a pairwise loss function to boost the performance. 
 User clicks are often used as training data since they can indicate relevance and are cheap to collect, but they contain substantial bias and noise. There has been some work on mitigating various types of bias in simulated user clicks to train effective learning-to-rank models based on multiple features. However, how to effectively use such methods on large-scale pre-trained models with real-world click data is unknown. To alleviate the data bias in the real world, we incorporate heuristic-based features, refine the ranking objective, add random negatives, and calibrate the propensity calculation in the pre-training stage. Then we fine-tune several pre-trained models and train an ensemble model to aggregate all the predictions from various pre-trained models with human-annotation data in the fine-tuning stage. Our approaches won 3rd place in the ``Pre-training for Web Search'' task in WSDM Cup 2023 and are 22.6\% better than the 4th-ranked team.  
  %要有逻辑，如果有次序，就第一，第二...（eg,先训基于不同策略训练了很多预训练模型，再集成...）
  % Ensemble methods were used at last and we won the $3^{rd}$ place at task2.
\end{abstract}

\keywords{Pre-training, Ensemble Learning, Negative Sampling}



\maketitle

\section{Introduction}
% background
It is essential to measure relevance effectively in various search-related scenarios, such as web search and e-commerce search. Standard approaches usually use human-annotated relevance labels as guidance to train a ranker. Although the relevance judgments between query-document pairs have high quality, they require extensive manual efforts and huge costs. State-of-the-art rankers are mainly based on pre-trained language models, which are of large scale and data-hungry. Training such models with limited manual relevance judgments may lead to sub-optimal performance. User clicks could be alternative or supplementary to training such large models since they are cheap to collect and can indicate relevance. 

% WSDM Cup 2023 presents a problem about how to use massive search engine data with bias to improve retrieval performance, so that users can get a better search experience. Among the two tasks in the competition, the purpose of "Pre-training for Web Search" task is to use real user search click logs (noisy and inaccurate)  and human-labeled dataset to pre-train and fine-tune the model respectively, and finally score the query-doc pairs in the online evaluation set to complete the ranking task. 

Click data is difficult to use effectively since it contains much noise and various of bias, including position bias, selection bias, trust bias, etc. Based on the large-scale search logs of a popular Chinese web search engine, the WSDM Cup 2023 presents a challenge to alleviate the bias in click data and use it for ranker training to improve retrieval performance. One of the two tasks in the competition is the ``Pre-training for Web Search'' task. It aims to train a BERT-style ranker with click data and the organizer also provide a human-annotated dataset an unbiased data source that can fine-tune the model. A hidden test set is used to evaluate each team's model performance at last. This task is a valuable foundation for academic studies on learning an unbiased pre-trained model for effective retrieval. 

Existing studies on unbiased learning mainly focus on learning-to-rank models, that typically learn an unbiased ranking function of various types of features such as relevance, recency, quality, and popularity using clicks as learning objectives. Semantic matching signals are often included as features in such models so the underlying model cannot be trained in an end-to-end manner. It is not clear whether existing unbiased learning methods can alleviate bias from click data and achieve good performance when training an end-to-end ranker based on a much larger pre-trained model. 

% The evaluation metric of the competition is DCG@10.
% add 0201
% Compared with unbiased learning to rank which only take the extracted features as the input, pre-training model can also process the text of the queries and the extracted document summary to capture the relationship between queries and documents via self-attention thus get better semantic representation. 

% challenges & our advantage
% 该句存疑，待考证
% However, there  lacks effective methods to deal with biased search logs with user feedback in information retrieval, existing work of pre-training models such as \cite{zou2021pre} mainly focus on the stage of pre-trained language model. Some methods of unbiased learning to rank such as Regression EM\cite{wang2018position} are specially designed for the task of learning to rank thus cannot be used in the pre-training task easily.
We investigate several representative unbiased learning methods in the task of pre-training for web search  and find that their benefit is not as large as in the learning-to-rank setting. We also observe that fine-tuning the BERT-style ranker with unbiased human-annotated data will improve the performance a lot. To mitigate multiple biases in the click data that harm model training and boost retrieval quality, we adopt multiple strategies in our runs, which are shown to be effective. Our solution has two stages: 1) pre-training the model with the click data, and 2) fine-tuning and ensemble based on the human-annotated data. In the first stage, we incorporate heuristic-based features, refine the ranking objective, add random negatives, and calibrate the propensity calculation. In the second stage, we augment training samples by duplicating high-frequency samples and learn an ensemble model using the predictions of our multiple runs and heuristics as features. We won 3rd place in the competition and our performance is 22.6\% higher than the team ranked 4th place.

% ours
% Considering the above challenges, we proposed effective strategies for the pre-training task. In this paper, we will introduce the techniques and methods we used in the competition. First of all, we hope to alleviate the bias problem in the click log by changing the training target in the pre-training stage. Specifically, we experimented with multiple solutions such as label correction, loss weighting, and long-tail data enhancement, expecting to obtain a better pre-training model. Secondly, we also tried some strategies to make better use of the expert annotation dataset, and finally we used LightGBM to integrate some characteristic model results as our final solution.

% \section{Methodology}
\section{Debiased pre-training}
In this section, we first describe the model architecture we use, and then we introduce our learning objective and inverse propensity weighting schemes.
% label correction, neg
\begin{figure*}
\centering
\includegraphics[scale=0.58]
% [width=
%         0.7\linewidth, height=0.25\textheight]
        {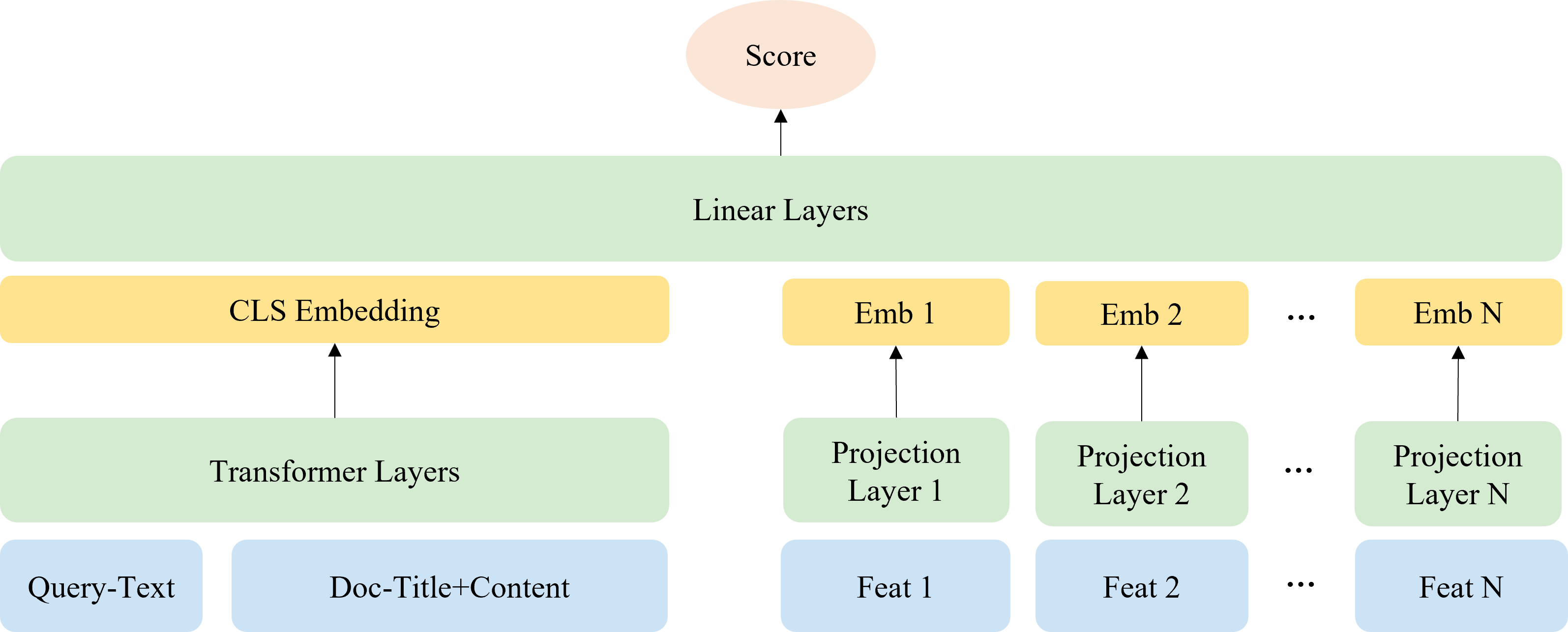}
% \caption{Pre-training Mode Architecture}
\caption{Model architecture.}
\label{pic:model}
\end{figure*} 
\subsection{Model Architecture}
% 璐璐
% 扩大维度的好处说法有点问题应该 等待被修改！- 感觉也可以不说
To leverage more matching signals, we include heuristic-based features such as BM25 \cite{robertson1994some} and query likelihood with multiple smoothing methods \cite{zhai2004study} in our model, which is as same as the procedure describe in \cite{yu2023feature}. As shown in Figure \ref{pic:model}, our model has a wide and deep architecture that learns semantic matching using deep neural networks and incorporates much more exact matching features (e.g., TF, IDF, TF-IDF, BM25, DIR \cite{zhai2004study}, etc.) with a shallow network. We adopt the well-known cross encoder, multi-layer bidirectional Transformer encoder blocks \cite{vaswani2017attention} (Hidden State=768, Attention Head=12, Layer=3 or 12), to encode the concatenated query and document text. Then the vector of the [CLS] token in the output of the encoder layers captures the interactions between the query and document terms. We project the heuristic-based dense features to hidden vectors and concatenate them with the embedding of [CLS]. Then MLP layers are used to map the concatenated vectors to a final score.  
% are concatenated together as the input of the Transformer encoder block, and the embedding vector of [CLS] token obtained from the model encoding is used as the cross semantic feature. Feed the combination of semantic features and pre-extracted features (after embedding) into the hidden layer DNN, and the output is the relevance score for this query-document pair.

 % Combine the semantic features and the statistical features extracted in advance as the input of the DNN with 512-256-128 hidden layers. As the number of the latter features is limited, we use the same projection matrix (3*8, 3 represents one feature computed using samples' title, content and the both ) to amplify their dimension, which makes the statistical features more discrete.
\begin{figure}[h]
\centering
\includegraphics[width=
        1.0\linewidth]{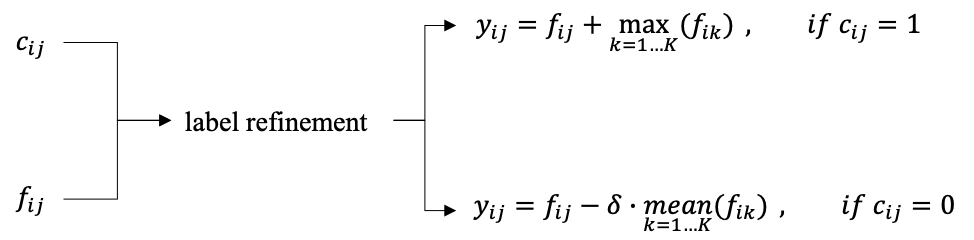}
\caption{The generation of the label $y_{ij}$ of each document in the corresponding list for each query $q_i$.}
\label{pic:labelt2}
\end{figure} 
\vspace{-0.2cm}
\subsection{Model Training}
% Add some new overall introduction about this subsection.
We introduce the strategies we use in model training, including label refinement and negative sample selection, and then we introduce the two loss functions we use.

\textbf{Label Refinement.}
We find that training an end-to-end ranker based on the pre-trained model on the click data, whether with or without unbiased learning, cannot outperform heuristic-based features such as BM25. So, instead of using clicks as labels, we obtain the learning targets by incorporating a well-performing heuristic-based feature with the clicks 
% Put the introduction about how we obtain the labels here. Information related to Figure 1. mention that how the heuristic label is obtained (multiple features aggregated together. ) 
according to Figure \ref{pic:labelt2}. In Figure \ref{pic:labelt2}, $c_{ij}$ represents whether the jth document in the document list given query $q_i$ is clicked, $f_{ij}$ is the corresponding feature value, and finally, we perform a Softmax operation on all $y_{ij}$ to get $\tilde{y_{ij}}$ so that all labels sum up to one.

\textbf{Negative Sample Selection.}
Only the top 10 documents are recorded in the click logs while the evaluation data includes the top 30 and other documents ranked in the top 900 with an interval of 30 (e.g., 30,60,90,...,990). Since top results are usually relevant and user clicks cannot cover all the relevant documents. Training a ranker with the click data will inevitably lead to sub-optimal performance on the test set. To let the model obtain the ability to discriminate relevant documents from irrelevant ones, besides the non-clicked documents in the top 10, we include random samples as negatives during training as well. 

\textbf{Loss Function.}
As previous studies show that listwise ranking loss has superior performance compared to pointwise and pairwise training, we train our model with a listwise loss based on attention allocation\cite{ai2018listwise}.
% [] (learning a deep listwise context model for ranking refinement). 
% introduce the loss function.
{\setlength\abovedisplayskip{1.1pt}
\setlength\belowdisplayskip{0.6pt}
\begin{equation}
\label{equ:list}
    L_{listwise} = -\sum_{i=1}^N\sum_{j=1}^K [\tilde{y_{ij}}\frac{exp(x_{ij})}{\sum_{k=1}^K exp(x_{ik})}]
\end{equation}}
The simple list-wise loss function is shown in the Equation \eqref{equ:list}, where $N$ is the total number of queries, $K$ is the length of the list, that is, the number of documents containing user feedback that is the longest saved in the click log. 
On this basis, since the documents displayed to the user already have different degrees of relevance, in order to make the model see more negative samples, we randomly add a specified number of random negative samples with different content of queries from the same training batch to the list, and set its label to zero. 
In addition, considering the documents after the last clicked one may not actually be observed by the user, we replace these documents with random negative samples to alleviate the false negative problem. 
% To let the model learn towards the heuristic-click-based labels outperform the heuristic itself, we also tried to relax the listwise matching constraint to pairwise so that the model has more freedom to behave differently from the target label. 
To encourage the model learned with the heuristic-click-based labels to outperform the original heuristic features, we also tried to relax the listwise matching constraint to pairwise so that the model has more freedom to behave differently from the target label. 
% introduce the principles to construct pairs. 
We attempt to build the order and use the pairwise loss function to optimize (see Section \ref{section:rank} for details). The priority order between two documents is mainly constructed based on the following relationships:

\begin{itemize}
\item Documents that are clicked are more relevant than documents that are not clicked.
\item Documents with high feature values are more relevant than documents with low values.
\item Documents shown to users outperform random negative samples.
\end{itemize}
Then the relative order can be obtained and we optimize the pre-training model using this objective.

\subsection{Inverse Propensity Weighting}
% DLA, fixed
%  璐璐+奕婷
Previous studies have shown the effectiveness of the Dual Learning Algorithm (DLA) in the task of unbiased learning to rank. However, we find that the propensity weights learned by DLA from the click data are not decreasing from the first to the tenth position, which is inconsistent with existing research. Also, since we conduct label refinement and its impact on propensity learning is unknown, we use static propensity calculated from click ratios for model training. The output of model variants trained with DLA and click-ratio-based propensities are used as features in the final ensemble model (in Section 3.4). The two inverse propensity weighting mechanisms are as follows:
%Though effectiveness, as we introduce the label correction strategy 
%- and the unbalanced sample re-weighting strategy 
%in this paper so the condition for DLA is different from ours. In order to alleviate such issue, we adopt two propensity weighting techniques as follows. 
\begin{itemize}
\item \textbf{DLA. } Using the dual learning algorithm, the propensity model and the ranking model are 
jointly optimized. It is worth mentioning that we fixed the propensity weight value of addictive negative samples to 1, which is slightly different from the original DLA. 
\item \textbf{Click-ratio-based propensity. } We use static inverse propensity weights according to the click ratio on different ranking positions of the whole training set in \cite{zou2022large} as the following: 
{\setlength\abovedisplayskip{1.1pt}
\setlength\belowdisplayskip{1pt}
\begin{equation}
    pw_i=(\frac{cr_1}{cr_i})^{\alpha},i=1,...,10.
\end{equation}}

Here, $\alpha$ is used to control the relative size of propensity weights and we fix it as 0.25. The fixed inverse propensity weights are set to 1, 1.19, 1.44, 1.58, 1.89, 1.95, 2.12, 2.26, and 2.51 for the top-10 ranked documents. 
\end{itemize}
\vspace{-0.3cm}

\section{Fine-tuning}
In this section, we describe the strategies we use in the fine-tuning stage including efficient design of ranking loss, negative sample selection, sample augmentation, and ensemble. 
\subsection{Ranking Loss}\label{section:rank}
% 介绍负样本选择以及loss设计（多标签样本）
% 晓洁
The document list under a query contains 5 classes: \{bad, fair, good, excellent, perfect\}. The most common way to deal with multi-level relevance labels is to transform them into positive and negative classes, thus using a point-wise loss function. 
% Common point-wise loss is defined as:
% \begin{equation}
%     L_{point-wise} = -\sum_{i=1}^N [y_{i}\times log(\sigma(x_{i}))+(1-y_{i})\times log(1-\sigma(x_{i}))]
% \end{equation}
% Where $\sigma$ is the sigmoid function, $N$ is the total number of q-d pairs selected from training data, $y_i$ is the binary label of example $i$.
Essentially, the point-wise method is to approximate the ranking problem to a regression problem, but the ranking task does not pursue accurate scoring, and relative scoring is acceptable. At the same time, the training of the model will be dominated by queries with a large number of labeled documents. Therefore, a pairwise method for modeling the relative relationship between positive and negative samples is needed. The pairwise method organizes a sample as $<q,d^+,d^->$, which means that $d^+$ is more relevant to $q$ than $d^-$.
{\setlength\abovedisplayskip{1.1pt}
\setlength\belowdisplayskip{1pt}
\begin{equation}
\label{pair1}
    L_{pairwise} = -\sum_{i=1}^N [\frac{exp(x_{i}^+)}{exp(x_{i}^+)+exp(x_{i}^-)}]
\end{equation}}

In order to improve the retrieval performance of the model, as shown in Equation \eqref{pair2}, we further introduce a certain number of negative samples into the pair-wise loss function formula, so that the model can better improve the relative score of positive samples. $T$ is the number of negative samples for each query plus 1, so a sample becomes $<q, d_{1}^-,d_{2}^-,..,d_{T-1}^-,d_T^+>$.
{\setlength\abovedisplayskip{1.1pt}
\setlength\belowdisplayskip{0.6pt}
\begin{equation}
\label{pair2}
    L = -\sum_{i=1}^N  [ \frac{exp(x_{iT}^+)}{\sum_j^T exp(x_{ij})}]
\end{equation}}

\subsection{Training Labels}
% 晓洁
Since the expert annotation dataset contains multi-level labels, we degenerate the 5 levels into two when using the pointwise and pairwise loss function due to its simplicity. We think that if a document is marked as "perfect" (label=4), "excellent" (label=3), or "good" (label=2), then the document is a positive sample, otherwise, the document is a negative sample. $d^+$ in Equations \eqref{pair1} and \eqref{pair2} is sampled from all positive samples of a query, and $d^-$ is selected from its negative samples. We also tried to train the model with the listwise loss according to Equation \eqref{equ:list} but got no gain. It is worth mentioning that probably we have not tuned the model sufficiently and more fine-grained exploration of the multi-grade label and more effective loss functions can be used to fine-tune in the future. 

% During specific training, there are two optional options. One is that different queries only appear once in an epoch, and their positive and negative samples are obtained by sampling. The other is that the number of positive samples determines the frequency of each query during training, and it is considered that queries with more positive samples should get more training opportunities.

\subsection{Sample Augmentation}
% 奕婷
Unlike click data in the search log, the cost of manual labeling is high and thus may lead to a limited number of queries in the expert annotated dataset. Therefore, it is promising to explore practical methods to enhance the effect of core samples in the fine-tuning stage. The queries in the human annotation data are de-duplicated and we find that there exist much more tail queries than head queries in the provided dataset. The numbers of high, mid, and low-frequency queries are 1092, 1820, and 1789 in the annotation dataset, respectively. Table \ref{tab:labels} illustrates the distribution of the relevance labels for different frequency queries. 
% 上边在表格里加了不同频率的label分布还有不同频率的query数量 解释原因上还要提一下
% In the fine-tuning stage, the human-annotation set is relatively small so it is necessary to develop effective methods to enhance the effect of the core samples. 
% We found that the validation set is constructed by the organizer and it is annotated by the employed annotators. 
In this paper, we consider high-frequency queries as the core samples in this stage due to the following reasons. 

It is known that head queries are easier so long tail queries are more important to differentiate the ability of a search engine. 
Positive documents contribute to model training a lot since most queries contain only a few perfect documents. We find that head queries have more positive documents since there are more relevant resources in the annotation set and previous user clicks can help search engines rank documents for such queries. We train the model more on the head queries to encourage the model to learn common matching patterns sufficiently. 
\vspace{-0.cm}
\begin{table}[h]
\setlength{\abovecaptionskip}{0.15cm}
\setlength{\belowcaptionskip}{-0.2cm}
  \caption{Distribution of relevance labels.}
  \label{tab:labels}
  \begin{tabular}{ccccc}
    \toprule
    &&\multicolumn{3}{c}{Ratio of label}\\
    \cline{3-5}
    Grade&Label&High&Mid&Low\\
    \midrule
    Perfect & 4& $0.49 \% $ & $0.15 \% $ & $0.02 \% $ \\
    Excellent & 3& $12.99 \% $ & $8.00 \% $ & $2.71 \% $ \\
    Good & 2& $35.06 \% $ & $31.32 \% $ & $21.33 \% $ \\
    Fair & 1& $15.96 \% $ & $9.40 \% $ & $5.16 \% $ \\
    Bad & 0& $35.50 \% $ & $51.13 \% $ & $70.78 \% $ \\
  \bottomrule
\end{tabular}
\end{table}

Considering the above factors, we amplified the effect of the high-frequency samples which is more important and accurately annotated through duplicating these samples and feeding them together with other samples during the fine-tuning stage. This strategy enhances the performance significantly and we leave the exploration of other augmentation strategies to the future.

\subsection{Ensemble Strategy}
% 奕婷
A simple but effective way is adopted in the ensemble stage. Taking the extracted features and the output scores of previous pre-trained and fine-tuned models as input, we train a Gradient Boosting Decision Tree model using Light Gradient Boosting Machine (LightGBM) \cite{ke2017lightgbm} in the validation set with LambdaRank objective, and the process can be formulated as follows: 
{\setlength\abovedisplayskip{1.1pt}
\setlength\belowdisplayskip{0.6pt}
\begin{equation}
    % score = argmin L(f(x); r)
    % \DeclareMathOperator*{\argmax}{arg\,max}
    % 没写完
    f^* = \mathop{\arg\min}_{f}{\mathbb{E}_{r} L (r, f(x))}
\end{equation}}

 The core hyper-parameters are num\_leaves, max\_depth, learning\_rate, and num\_iteration. We tried different combinations of these parameters and we tuned the hyper-parameters on a small proportion of the validation set, which is described in Section 4. We tried normalization to the input but got no gain.  

\section{Experiments}
% \subsection{Settings}
% yiting
% delta, tao怎么写 待加入
 % In order to utilize the biased data effectively, 
 Considering that queries with fewer than 10 retrieved documents may be too specific and not likely to have truly relevant results, we filter out the query-document pair with no clicks and those queries with less than 10 candidate documents. We use the AdamW optimizer. 
 % The value of $\beta$ and $\gamma$ of the unbalanced sample re-weighting technique is set as 1.7 and 0.75 respectively. 
 We use $\delta=2$ and $\tau=0.1$ in the stage of label refinement. Our method is implemented in PaddlePaddle and we use the official model for initialization. In the fine-tuning stage, the provided validation set (i.e. human-annotation set) is randomly divided into two parts using the query-id, and we use 80\% for training and 20\% for validating. We select the best model based on the performance of the 20\% validation set. 
% \subsection{Results}
% 比赛排名+集成之前各个模型的分数，集成后的分数
% finetune的分数不太好看，不想放集成的了
% 晓洁
Table \ref{tab:res} shows the top 5 teams and their final scores in this task. Our solution got $9.83$ for the online dcg@10 and won 3rd place.

\begin{table}[h]
  \setlength{\abovecaptionskip}{0.1cm}
\setlength{\belowcaptionskip}{-0.1cm}
  \caption{Top 5 scores of the competition. Our team won 3rd in the public score leaderboard.}
  \label{tab:res}
  \begin{tabular}{ccl}
    \toprule
    Rank & Team Name & DCG@10\\
    \midrule
    1 & Tencent Search & $12.16525  $ \\
    2 & THUIR & $10.04097  $ \\
    \textbf{3} & \textbf{Cannot Retrieve} & \textbf{9.83148} \\
    4 & Accepted & $8.02173  $ \\
    5 & DisTime & $7.30951  $ \\
  \bottomrule
\end{tabular}
\end{table}
\vspace{-0.4cm}

\section{Conclusion}
In this paper, we detail our winning solution to the task of pre-training for web search in WSDM Cup 2023. We use the Transformer model as the backbone, with the combination of negative sample augmentation and target modification
% , unbalanced sample re-weighting 
in the pre-training stage. A pairwise ranking loss, key sample augmentation, and ensemble strategy are used in the fine-tuning stage. Our solution achieved the DCG@10 score of 9.83148 and finally, we ranked 3rd on the leaderboard.
\begin{acks}
% Thanks to all the organizers and sponsors of WSDM Cup 2023, as well as the Baidu search team who provided the "Large Scale Web Search Session Data" dataset.
This work was supported by the Lenovo-CAS Joint Lab Youth Scientist Project. Any opinions, findings, conclusions, and recommendations expressed in this material are those of the authors and do not necessarily reflect those of the sponsors.
\end{acks}

\bibliographystyle{ACM-Reference-Format}
\bibliography{sample-base}


\end{document}